\documentstyle[11pt,epsfig,psfig]{article}
\begin{document}
\title{BES R measurements and $J/\psi$ decays }


\author{{Ning Wu}\\
{\small Division 1, Institute of High Energy Physics, P.O.Box 918-1,
Beijing 100039, P.R.China}}
\date{}
\maketitle

\vskip -0.5cm

\noindent

R measurement results in 2 -- 5 GeV region by BESII is 
reported in this talk. The study on $\sigma$ particle
in $J/\psi \to \omega \pi^+ \pi^-$, based on 
$7.8 \times 10^6$ BESI $J/\psi$  data, is also reported.




\section{BES R Measurements}

\begin{figure}[htbp]
\centerline{\epsfig{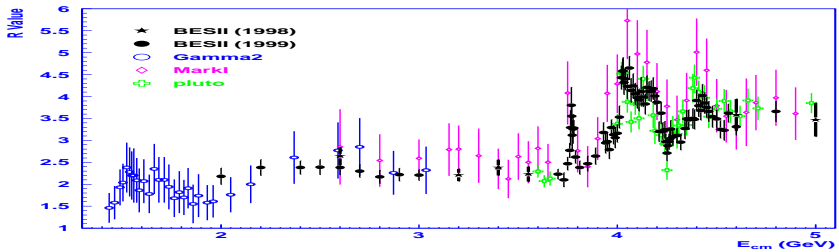}}
\vspace{-0.2in}
\caption[]{ R values in 1 -- 5 GeV energy region }
\label{r01} 
\end{figure}

R is defined as,
$$
R = \frac{\sigma(e^+ e^- \to ~ hadrons)}
{\sigma(e^+ e^- \to ~ \mu^+ \mu^-)}
= 3 \sum_q Q_q^2
$$
A precision measurement of R will reduce the uncertainty 
on $\alpha(M_z)$, narrow the predicted mass window of 
Higgs boson $m_H$ and reduce the uncertainty on 
the anomalous magnetic moment of muon $a_{\mu}$. 
R has been measured in the energy region from hadron 
production threshold to $Z^0$ pole. Previous R measurements
below 5 Gev were carried out almost 20 years ago and the 
largest uncertainty on R comes from the 
measurement below 5 GeV energy region. 
Therefore, it is important to measure R value in the 
2 -- 5 GeV energy region. 
BES collaboration has carried out two runs of R scan. 6 energy
points were scanned in the 1998's run. Data samples at 2.6 GeV
and 3.55 GeV were used to tune LUND parameters\cite{1}. 
In the second run of 1999, 
85 energy points were scanned, from which 24 points with
separated-beam data and 7 points with sigle-beam data. 
In the experiment, R is determined by the following realtion:
$$
R = \frac{\sigma^0_{}had}{\sigma^0_{\mu \mu}}
= \frac{N^{obs}_{had} - N_{bg} - \sum_l N_{ll} - N_{\gamma \gamma}}
{\sigma^0_{\mu \mu} \cdot \varepsilon_{had} \cdot 
\varepsilon_{trg} \cdot (1 + \delta) \cdot {\cal L} },
$$
where $N^{obs}_{had}$ represents the observed hadronic events, 
$N_{bg}$ represents the beam-associated backgrounds, 
$N_{ll}$ represents the lepton pair backgrounds,
$N_{\gamma \gamma}$ represents two photon process events,
$\varepsilon_{had}$ stands for the detection efficiency,
$\varepsilon_{trg} $ is the trigger efficiency,
$(1 + \delta)$ is the radiative correction,
${\cal L}$ is the integrated luminosity 
and $\sigma^0_{\mu \mu}$ is the lowest cross section
of $e^+e^- \to \mu^+ \mu^-$ which is given by
$\sigma^0_{\mu \mu} (s) = 4 \pi \alpha^3 /3 s$.
One of the most important work in R measurements is hadronic
event selection. It is done in three steps. First remove 
apparent Bhabha events. Then in track level, select track to
remove cosmic rays, beam-gas. Finally, in event level, remove
beam-gas, two-photon, $e e$, $\mu \mu$, $\cdots$.
The main backgrounds come from cosmic rays, lepton pair
production, two-photon processes and beam associated background.
The most serious background is beam associated background.
It is studied in two independent methods. The first method is
to use separated-beam data to calculate the event number of
beam associated background, the second method is to use z 
component of event vertex to calculate the event number.
The results given by two different method are consistent. 
Detection efficiency is given by Monte Carlo simulation,
which is model dependent. However, results given by different
models are consistent in 3 percent level. 
The integrated luminosity is determined by
$
L = \frac{N_{obs}}
{\sigma \cdot \varepsilon \cdot \varepsilon_{trg}},
$
with $N_{obs}$ the number of events (Bhabha, dimuon and 
$\gamma \gamma$), $\sigma$ the production cross section,
$\varepsilon_{trg}$ the trigger efficiency and
$\varepsilon$ the detection efficiency.
Initial state radiative correction is studied in four
different schemes and they are consistent within 
$1 \%$ in continuum region. Finally, Crystall Ball scheme
is selected to calculate the radiative correction. 
Uncertainties of R values are reduced down to $ 6 \sim 10 \%$. The
average uncertainties of R is about $6.6 \%$. As an example, 
the error source of R value at 3.0 GeV energy point is
shown in Table 1. R values in 1 -- 5 GeV energy region
is shown in Fig.\ref{r01}\cite{2}.

\vspace{-0.1in}
\begin{table}[htp]
\begin{center}
\doublerulesep 0pt
\renewcommand\arraystretch{1.1}
\begin{tabular}{|l|l l l l|l l|l|}
\hline
\hline
\hline

Source      &	$N_{had}$  &L  &  $1+ \delta$  & 
$\varepsilon_{had} $ & Sys.  &  Stat.  & Total   \\
\hline

Error contribution ($\%$) &  3.3 &  2.3 &  1.3&  3.0
&  5.2&  2.5 &  5.8  \\

\hline                                                 
\hline                                                 
\hline                                                 
\end {tabular}                                         
\caption { Error Source (3.0 GeV as example) }
\end{center}
\end{table} 

\vspace{-1.0in}

\section{Search  for $\sigma$ in $J/\psi$ Hadronic Decay}

The early analysis of $\pi \pi$ and $\pi K$ scattering data
shows no pole at the lower mass region. Up to now,  
the existence of $\sigma$ and $\kappa$  as  resonant 
particles has not been widely accepted. However, recent
re-analysis of the $\pi \pi$ and $\pi K$ scattering data
shows an evidence for existence of the $\sigma$ and 
$\kappa$ particle with comparatively light mass\cite{3}. 
If $\sigma$ is a s-channel resonance, it should appear
in production process. Therefore, it is important to 
search for $\sigma$ and $\kappa$ in  production 
process. \\

{\bf
\begin{figure}
\vspace {-0.25in}
\begin{flushleft}
{\mbox{\psfig{file=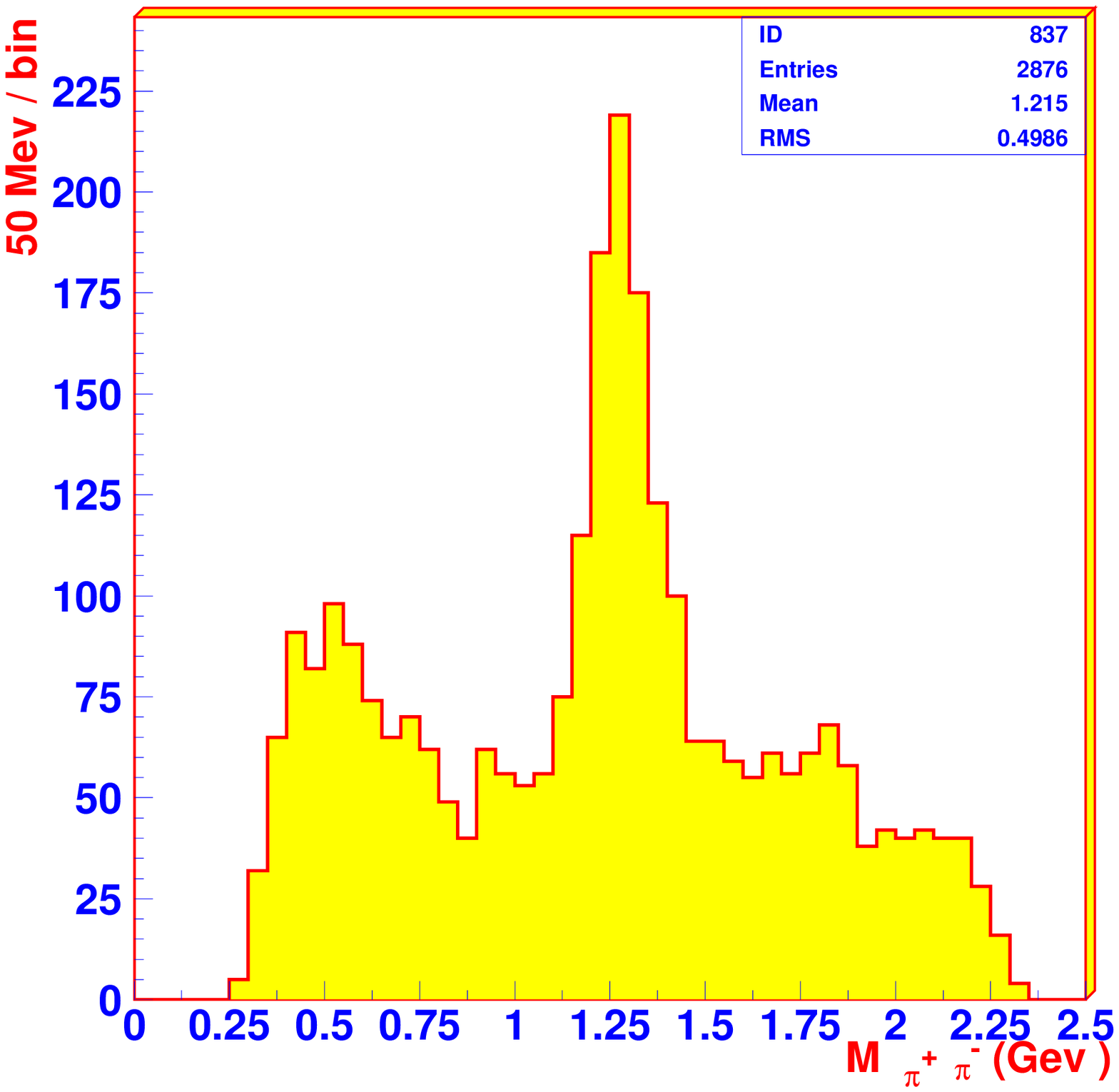,height=1.7in,width=1.7in}}}
\end{flushleft}
\vspace {-2.05in} 
\begin{center}   
{\mbox{\psfig{file=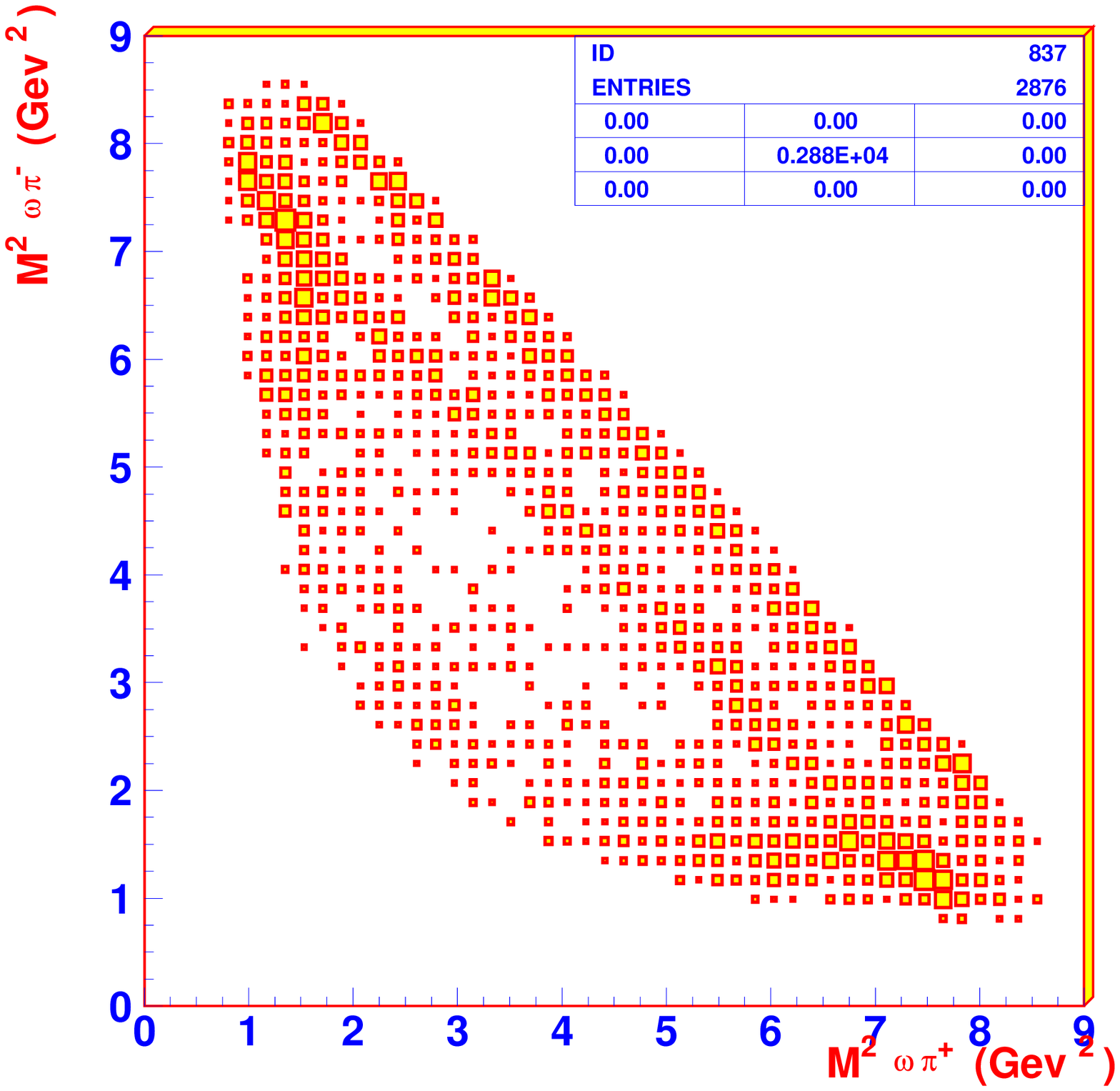,height=1.7in,width=1.7in}}}
\end{center}     
\vspace {-2.05in}
\begin{flushright}
{\mbox{\psfig{file=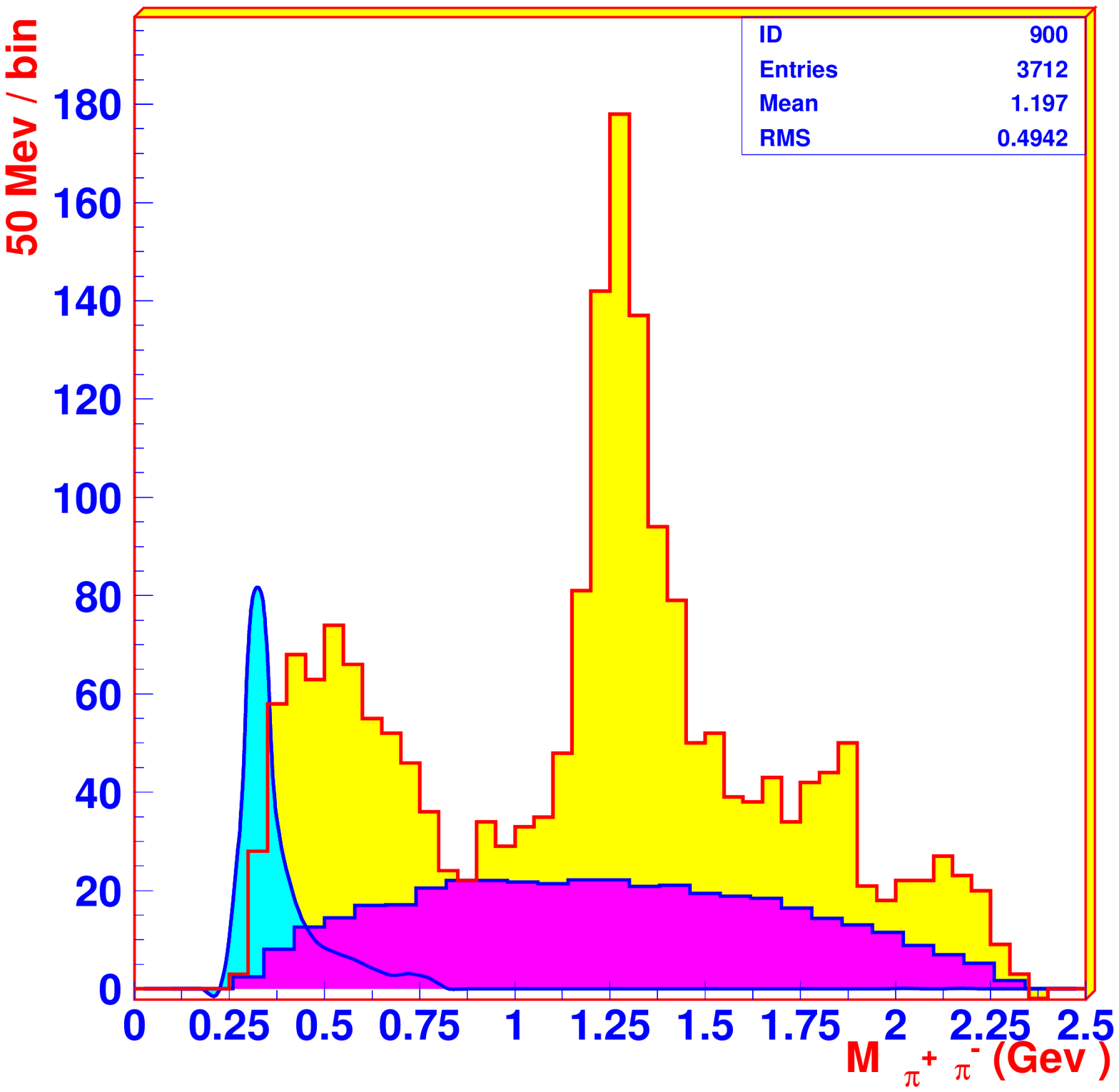,height=1.7in,width=1.7in}}} 
\end{flushright}
\vspace {-0.3in}
\caption[]{ LEFT:The invariant mass spectrum of $\pi^+ \pi^-$;
MIDDLE:Daliz plot; RIGHT: mass spectrum after side-band
subtraction.  }
\label{w01} 
\end{figure}
}

One found that there is a low mass enchancement 
in the $\pi^+ \pi^-$ invariant mass spectrum in
$J/\psi \to \omega \pi \pi$ (Fig.\ref{w01}).
For the sake of complicity, we call this low
mass enhancement the $\sigma$-particle.
In Daliz plot, a clear band which corresponds to 
the $\sigma$-particle can be seen(Fig.\ref{w01}).
There are some backgrounds in the above data sample
(such as $J/\psi \to \rho 3 \pi$, $\cdots$). Some 
of the backgrounds are shown in the $\omega$ side-band, 
which can be removed through side-band
subtraction. $\sigma$-particle can still be observed
in the $\pi^+ \pi^-$ invariant mass spectrum
after side-band subtraction
(Fig.\ref{w01}). Threshold effects and phase space
effects can also be seen in this figure.
Apparently, the first peak does not originate 
from threshold effect and phase space effect.
We also performed detailed Monte Carlo study and found
that the background events from other $J/\psi$
decay channels are very less. Therefore, the first
peak is  likely a s-channel resonance. 
Then, a Partial Wave Analysis(PWA) on
the $\pi^+ \pi^-$ invariant mass spectrum 
is applied to
study the structure of the first peak.
Our results strongly favor that
the spin-parity is $0^{++}$ and the statistical
significance of $\sigma$ is about 18 $\sigma$.
Its mass and width are determined through mass
and width scan and the results are listed in Table 2. 
Our scan curve is shown in Fig.\ref{w02} with the cross
representing real data and the histogram
 being fit curve. The final fit
on the global mass spectrum is also shown
in Fig.\ref{w02}. $f_0(980)$ also appears in this 
channel. Our results on $f_0(980)$ is  listed in
Table 2.  The branching ratios listed there are only 
for $J/\psi \to \omega X \to \omega \pi^+ \pi^-$.

{\bf
\begin{figure}   
\vspace {-0.25in} 
\begin{flushleft}
{\mbox{\psfig{file=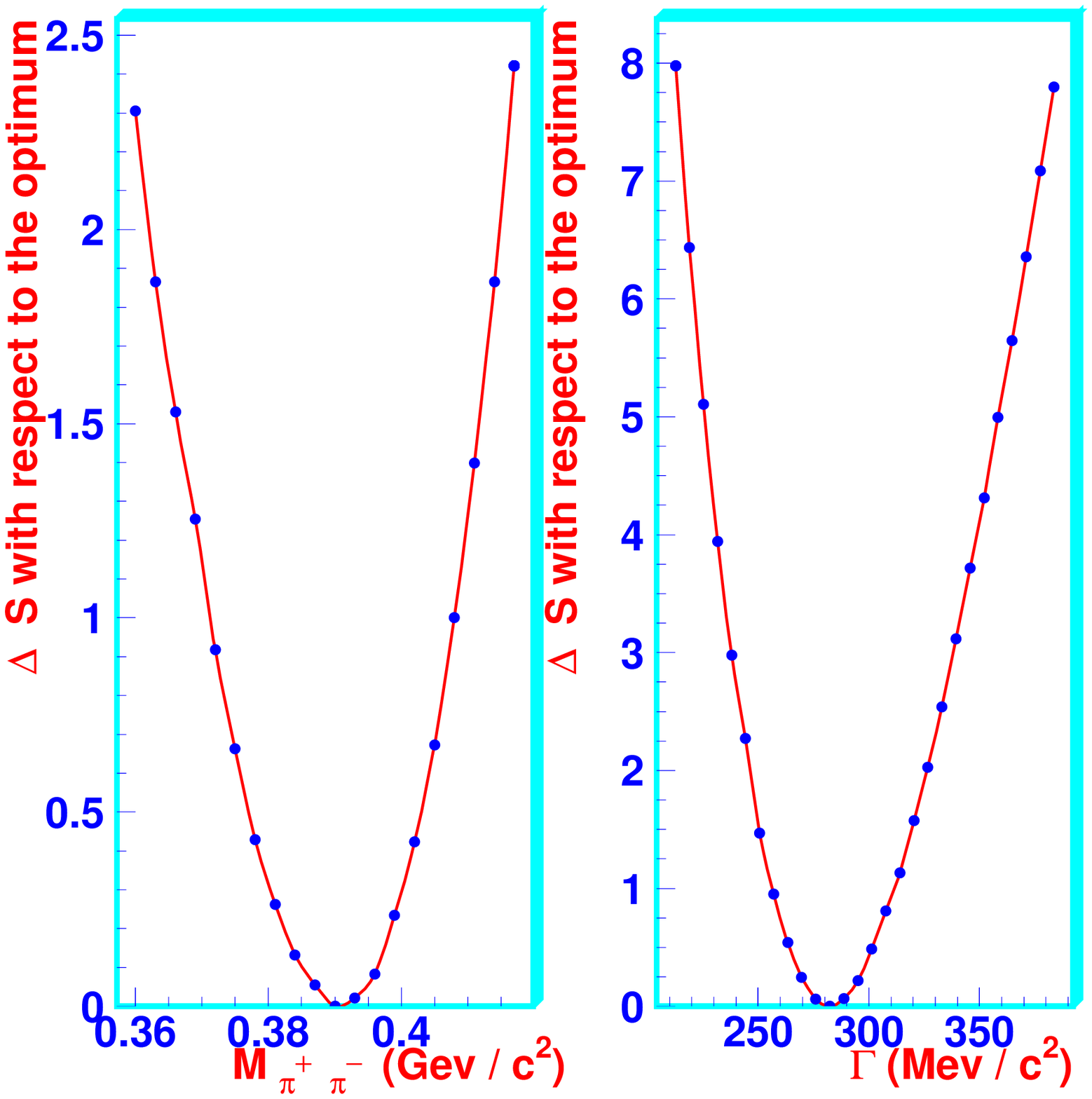,height=1.7in,width=3.5in}}}
\end{flushleft}  
\vspace {-2.05in}
\begin{flushright}
{\mbox{\psfig{file=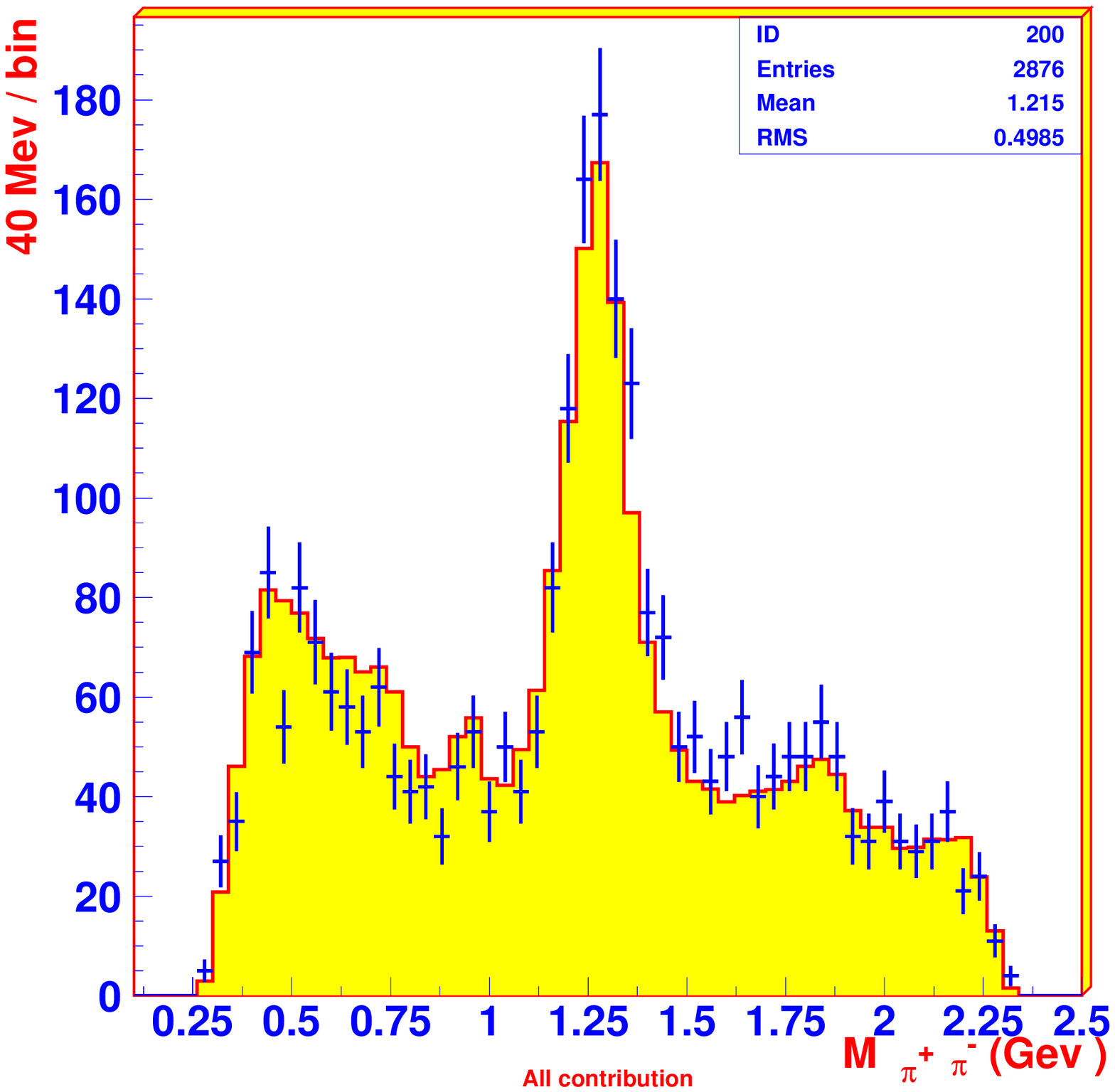,height=1.7in,width=1.7in}}}
\end{flushright}
\vspace {-0.3in}
\caption[]{ LEFT: Mass scan on $\sigma$; MIDDLE:width scan 
on $\sigma$; RIGHT: final global fit(error bar is real data 
and histogram is fit ).  }
\label{w02} 
\end{figure}
}

\vspace {-0.27in}
{\normalsize
\begin{table}[htp]
\begin{center}
\doublerulesep 0pt
\renewcommand\arraystretch{1.0}
\begin{tabular}{|l|l|l|l|}
\hline
\hline
\hline

Resonance  &  M  (MeV)  & $\Gamma$  (MeV)  &  
                BR ($\times 10^{-4}$) \\ 
\hline

$\sigma$-particle &  $390^{+60}_{-36}$  &
        $282^{+77}_{-50}$  &  $17.1 \pm 3.4 \pm 4.3$   \\
\hline

$f_0(980)$ &  $976^{+22}_{-20}$  &
        $78^{+63}_{-43}$  &  $2.80 \pm 0.56 \pm 0.4$ \\
\hline

$f_2(1270)$ &  $1280^{+13}_{-12}$  &
        $164^{+29}_{-27}$  &  $43.5 \pm 8.7 \pm 5 $ \\
\hline  

\hline
\hline
\end {tabular}
\caption { Final results of some main resonances.}
\end{center}
\end{table} 
}

\vspace {-0.2in}

$\sigma$, as a s-channel resonance, 
 is not found in the 
$J/\psi \to \gamma \pi^+ \pi^-$
It means that it does not look like a scalar glueball. 
According to quark model, the ordinary $q \bar{q}$ 
scalar  meson nonet with lowest mass is
$1 ^3P_0$ states. Its orbital angular momentum 
is excited to P wave. The mass of them can not be as low
as 390 MeV. $\sigma$ can not be filled into the quark
model for ordinary $q \bar{q}$ mesons.

\end{document}